\documentclass[reprint,superscriptaddress,nofootinbib,amsmath,amssymb,aps,prl]{revtex4-1}

\usepackage{amsmath,amssymb}
\usepackage{amsthm}
\usepackage{mathtools}
\usepackage[noend]{algorithmic}
\usepackage[ruled,vlined]{algorithm2e}
\usepackage{setspace}
\usepackage{url}
\usepackage{makeidx}
\usepackage{enumerate}
\usepackage{graphicx,float,psfrag,epsfig}
\usepackage[usenames,dvipsnames,svgnames,table]{xcolor}
\definecolor{darkgreen}{rgb}{0.0,0,0.9}
\usepackage[pagebackref,colorlinks=true,pdfpagemode=UseNone,citecolor=OliveGreen,linkcolor=BrickRed,urlcolor=BrickRed,pdfstartview=FitH]{hyperref}
\usepackage{epstopdf}
\usepackage{color}
\usepackage{xr}
\usepackage{subfig}
\usepackage{graphicx}
\usepackage[utf8]{inputenc}
\usepackage{bm}

\usepackage{scalerel,stackengine}

\stackMath
\newcommand\reallywidehat[1]{%
\savestack{\tmpbox}{\stretchto{%
  \scaleto{%
    \scalerel*[\widthof{\ensuremath{#1}}]{\kern.1pt\mathchar"0362\kern.1pt}%
    {\rule{0ex}{\textheight}}
  }{\textheight}%
}{2.4ex}}%
\stackon[-6.9pt]{#1}{\tmpbox}%
}
\DeclareSymbolFont{rsfs}{U}{rsfs}{m}{n}
\DeclareSymbolFontAlphabet{\mathscrsfs}{rsfs}

\numberwithin{equation}{section}

\newtheoremstyle{myexample} 
    {\topsep}                    
    {\topsep}                    
    {\rm }                   
    {}                           
    {\bf }                   
    {.}                          
    {.5em}                       
    {}  

\newtheoremstyle{myremark} 
    {\topsep}                    
    {\topsep}                    
    {\rm}                        
    {}                           
    {\bf}                        
    {.}                          
    {.5em}                       
    {}  

\theoremstyle{myremark}

\theoremstyle{myremark}

\theoremstyle{myexample}

\definecolor{darkgreen}{rgb}{0.0, 0.5, 0.0}

\newcommand{\bea}{\begin{eqnarray}}
\newcommand{\eea}{\end{eqnarray}}
\newcommand{\<}{\langle}
\renewcommand{\>}{\rangle}
\newcommand{\E}{{\mathbb E}}

\def\fr{\frac}
\def\fr12{\frac{1}{2}}

\def\Unif{{\sf Unif}}

\def\id{{\boldsymbol{I}}}
\def\bh{\boldsymbol{h}}

\def\bP{\boldsymbol{P}}

\def\bP{{\boldsymbol{P}}}

\def\bh{{\boldsymbol{{h}}}}

\def\bx{{\boldsymbol{x}}}

\def\de{{\rm d}}

\def\E{{\mathbb E}}

\def\<{\langle}
\def\>{\rangle}

\def\by{{\boldsymbol{y}}}

\def\b0{{\boldsymbol{0}}}

\def\br{{\boldsymbol r}}

\usepackage{graphicx,subcaption}
\usepackage{xcolor}
\usepackage{amssymb}
\usepackage{amsmath}
\usepackage{dcolumn}
\usepackage{dsfont}
\usepackage{amsthm, amsfonts}
\usepackage{bbm}
\usepackage{comment}
\usepackage{eucal}

\begin{document}
\title{Learning limit cycles via Hebbian synaptic plasticity}

\author{Samantha J. Fournier}
\affiliation{Université Paris-Saclay, CNRS, CEA, Institut de physique théorique, 91191, Gif-sur-Yvette, France}
\author{Luca Vincenzo Spallanzani} 
\affiliation{Université Paris-Saclay, CNRS, CEA, Institut de physique théorique, 91191, Gif-sur-Yvette, France}
\author{Pierfrancesco Urbani}
\affiliation{Université Paris-Saclay, CNRS, CEA, Institut de physique théorique, 91191, Gif-sur-Yvette, France}

\begin{abstract}
We investigate high-dimensional, non-linear dynamical systems when exposed to incoherent periodic inputs and Hebbian-like synaptic plasticity. Our findings reveal a striking phenomenon: depending on the interplay between the strength of the periodic drive and synaptic plasticity, the system’s phase diagram can give rise to a region where, once both inputs are removed, the collective dynamics spontaneously settles into a limit cycle. This suggests that periodic drives can imprint lasting rhythmic patterns into the network through plasticity, effectively teaching it to oscillate on its own. Numerical simulations on finite size systems show that the limit cycle phase can be easily detected on single-sample trajectories, while averaged curves are affected by strong finite size effects due to sample-to-sample fluctuations of the period of the limit cycles.
\end{abstract}

\maketitle

\paragraph*{\bf Introduction --}
Oscillatory activity is a ubiquitous feature of brain dynamics and can emerge through several mechanisms, including recurrent neuronal interactions that give rise to synchronized population activity~\cite{Tsodyks2000Synchrony}, communication between distinct brain regions~\cite{Clark2025Multiregion}, and entrainment by an external periodic stimulus~\cite{rajan2010stimulus}. While these mechanisms describe how rhythmic dynamics can arise given a network's existing connectivity, a complementary and less understood question is how such dynamics can be acquired in the first place. Recurrent neural networks (RNNs) can be trained to autonomously generate arbitrary periodic patterns~\cite{sussillo2009generating}, a capacity thought to underlie functions such as motor control, memory consolidation, and sensory processing. Yet, how biological neural circuits might learn such rhythmic patterns through local synaptic plasticity, potentially guided by entrainment to an external periodic stimulus, remains a fundamental open question.

From a theoretical standpoint, the problem is twofold. First, while limit cycles—stable, self-sustaining oscillations—are a common feature of low-dimensional non-linear dynamical systems, they become increasingly difficult to obtain as dimensionality grows, where chaotic attractors become the more generic outcome. This is directly relevant for biological neural networks which operate in high-dimensions. For this reason, traditional models of learning have primarily focused on fixed-point and chaotic attractors, leaving the mechanisms underlying rhythm learning underexplored. 
Second, the interplay between external periodic inputs and synaptic plasticity creates a self-consistent feedback loop: entrained collective dynamics reshape synaptic connectivity, which in turn determines whether the resulting limit cycle remains stable after the periodic drive is removed. This non-linear, out-of-equilibrium coupling between neuronal activity and synaptic adaptation makes analytical treatment challenging.

In this work, we investigate whether limit cycle activity, transiently induced by external periodic stimulation, can be stabilized through synaptic plasticity. We make analytical progress by studying prototypical high-dimensional non-linear dynamical systems whose behavior mirrors that of RNNs in computational neuroscience. We focus on the case where such systems are driven by incoherent periodic inputs and subject to Hebbian-like plasticity. Our analysis reveals a novel regime in the phase diagram: one in which the combined effect of the periodic drive and plasticity imprints a memory of the rhythm onto the network's connectivity. Once the drive and plasticity are removed, the network relaxes onto a limit cycle, demonstrating that rhythmic patterns can be learned and self-sustained. This result sheds light on the emergence of oscillations in adaptive systems, and establishes a concrete link between driven and autonomously generated dynamics.

\begin{figure*}[t]
    \centering
    \includegraphics[width=1\textwidth]{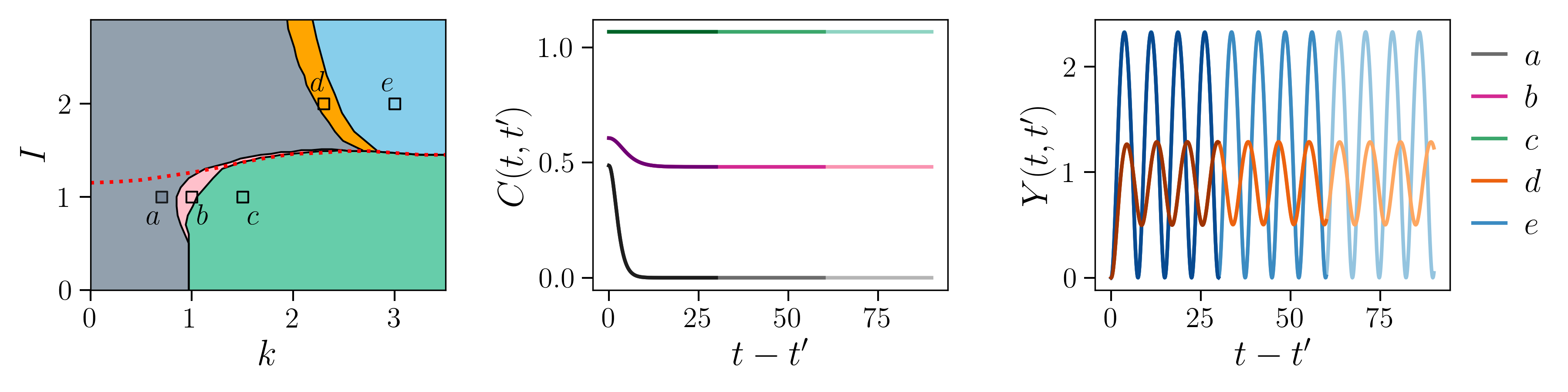}
    \caption{Phenomenology of steady-states in the memorization phase $t\gg T_\mathrm{ht}$, for $T_\mathrm{st}=T_\mathrm{ht}\gg 1$. (Left) Phase diagram with $a.$ Plain chaos (grey), $b.$ Semi-freezable chaos (pink), $c.$ Freezable chaos (green), $d.$ Semi-freezable limit cycle (orange) and $e.$ Freezable limit cycle (blue). The red dotted line denotes the transition from chaotic dynamics to limit cycle activity during the periodic stimulation ($t<T_\mathrm{st}$). (Middle and Right) Steady-state behavior of $C(t,t')$ and $Y(t,t')$ in the memorization phase ($t\geq t'\gg T_\mathrm{ht}$). Color shadings correspond to different values of $t'$.}
    \label{fig:phase_diagram}
\end{figure*}

\paragraph*{\bf Related literature --}
The study of high-dimensional, non-linear dynamical systems for biological neural networks has deep roots in statistical mechanics, with foundational contributions by Sompolinsky, Crisanti, and Sommers (SCS) who established the theoretical framework for RNNs in the thermodynamic limit~\cite{Sompolinsky1988chaos}. A key result from this line of work is the absence of limit cycles in the phase diagram of such systems: depending on the parameters, the dynamics either converge to a fixed point or exhibit chaotic behavior—but stable periodic orbits do not emerge spontaneously.
The problem of learning limit cycles has been explicitly addressed in computational neuroscience, most notably by Sussillo and Abbott, who introduced the FORCE (First-Order Reduced and Controlled Error) learning algorithm to train RNNs to generate periodic attractors~\cite{sussillo2009generating}. While FORCE demonstrates the feasibility of learning rhythmic patterns (see also~\cite{laje2013robust, rajan2016recurrent, Nicola2017FORCE, depasquale2018full}), the precise algorithmic strategy employed uses biologically implausible plasticity rules~\cite{asabuki2025predictivealignment}.

The response of random neural networks to external drives has been extensively investigated. Rajan, Abbott, and Sompolinsky analyzed the impact of incoherent periodic inputs on RNN dynamics, revealing how such drives can shape collective oscillations~\cite{rajan2010stimulus}. Complementarily, the role of synaptic plasticity in shaping network dynamics was explored by Clark and Abbott, who demonstrated how adaptive synapses can induce transitions from chaotic to fixed-point regimes~\cite{clark2024theory}. Our work builds on these insights but adopts a more abstract, prototypical approach. We consider non-linear dynamical systems that, despite their simplicity, capture the essential phenomenology of more biologically detailed models. In particular, the class of models considered in this work has been shown to reproduce the key observations of Rajan et al.~and Clark and Abbott~\cite{fournier2023statistical, fournier2025non, fournier2026high, fournier2026chaos}.
We extend these works to reveal a novel mechanism: the learning of limit cycles through the interplay of incoherent periodic drives and Hebbian-like plasticity. This bridges the gap between driven and autonomous dynamics, offering a theoretically tractable yet biologically inspired perspective on rhythm learning.

\paragraph*{\bf The model --}
We consider high-dimensional, non-linear dynamical systems previously studied in~\cite{fournier2023statistical, fournier2025non, fournier2026high, fournier2026chaos, urbani2026theory}.
{Unlike models that aim to capture microscopic neuronal mechanisms (e.g.~the model of SCS~\cite{Sompolinsky1988chaos}), these systems are motivated by the observation that biologically inspired RNNs are high-dimensional, non-linear systems whose random recurrent interactions give rise to sustained chaotic fluctuations that can be exploited for computation~\cite{generativeRNN}}.
Thus, we focus on systems that are high-dimensional, non-linear, and include random interactions to generate chaotic activity.

Consider a $N$-dimensional state vector $\bx(t)$ governed by
\begin{equation}
\begin{split}
    \dot \bx(t) &= -\mu(t)\bx(t) + g\, \br(\bx(t)) + \bP(t) + \bh(t)\\
    \bx(0) &\sim \mathcal{N}(0, \id).
\end{split}
    \label{dyn_sys}
\end{equation}
The right-hand side of Eq.~\eqref{dyn_sys} has four terms. The first, $-\mu(t)\bx(t)$, confines the dynamics near the origin. We assume $\mu(t) = \hat{\mu}(C(t,t))$, where $C(t,t) = |\bx(t)|^2 / N$, and $\hat{\mu}(z)$ grows sufficiently fast to confine the system.

The second term, $g\, \br(\bx(t))$, provides endogenous chaotic fluctuations. The random Gaussian field $\br(t)$ satisfies the following statistics
\begin{equation}
    \begin{split}
        \mathbb{E}[\br(\bx)] &= 0 \\
        \mathbb{E}[r_i(\bx)r_j(\by)] &= \delta_{ij}\, G\left(\frac{\bx \cdot \by}{N}\right).
    \end{split}
\end{equation}
For non-linear $G(z)$, the system is non-linear. We use $G(z) = g_1^2 z + 2g_2^2 z^2$, but our analysis applies to any $G(z)$ with positive Taylor coefficients. The parameter $g$ sets the strength of the chaotic drive. Confining and chaotic drive terms alone can yield chaotic or fixed-point phases, with chaos emerging for sufficiently strong, non-linear $\br$ \cite{fournier2026high}.

The Hebbian term $\bP$ introduces plastic interactions, inspired by~\cite{clark2024theory}
\begin{equation}
    P_i(t) = \sum_{j=1}^N A_{ij}(t) x_j(t).
\end{equation}
The plastic matrix $A_{ij}(t)$ evolves as
\begin{equation}
\begin{split}
    p\, \dot{A}_{ij}(t) &= -A_{ij}(t) + \frac{k}{N}\, x_i(t) x_j(t) \\
    A_{ij}(0) &= 0.
\end{split}
\end{equation}
Here, $p$ sets the plasticity timescale, and $k$ sets its amplitude, following Hebbian theory.

We study the system driven by an external periodic input
\begin{equation}
    h_i(t) = I\, \cos(\omega t + \theta_i)
\end{equation}
with $\theta_i \sim \text{Unif}(0, 2\pi)$ modeling random phase shifts. The frequency $\omega$ and amplitude $I$ are constant parameters.

Finally, adapting the terms in Eq.~\eqref{dyn_sys}, allows modeling the SCS system~\cite{Sompolinsky1988chaos} by setting $\mu(t) = 1$, $r_i(\bx(t)) = \sum_j J_{ij} \phi(x_j)$, and $P_i(t) = \sum_j A_{ij}(t) \phi_j(t)$, with $\phi$ a non-linear activation function. The standard choice that we adopt is $\phi=\tanh$.

\paragraph*{ \bf The learning protocol --}
We consider a two-stage learning setting.

\paragraph*{Stimulation phase: $t\in [0, T_\mathrm{st}]$ --} The dynamics of the system starts at $t=0$ and receives the periodic input $\bh$ for a long stimulation time $T_\mathrm{st}$. Correspondingly, the matrix $A$ changes with time due to the plasticity rule. 
We assume that the stopping time $T_\mathrm{st}$ at which the periodic input is removed is sufficiently long for the dynamical system to reach a stationary state that we characterize. 

\paragraph*{Memorization phase: $t>T_\mathrm{ht}$ --}
At $t=T_\mathrm{ht}\geq T_\mathrm{st}$ the synaptic plasticity is switched off. 
Correspondingly, for times larger than the halting time, $t>T_\mathrm{ht}$, one has $A_{ij}(t)=A_{ij}(T_\mathrm{ht})$ and $\bh(t)=0$.
We are interested in understanding to what extent the dynamics after the halting of plasticity and periodic input is locked into a limit cycle.

\paragraph*{Metabolic phase: $T_\mathrm{st}< T_\mathrm{ht}$ --} In general, $T_\mathrm{ht}$ and $T_\mathrm{st}$ are different: the synchronicity between plasticity and input in biological contexts is possibly driven by chemicals with a finite relaxation time. While we will analyze the limiting case in which $T_\mathrm{ht}=T_\mathrm{st}$, we will show that, as far as a $T_\mathrm{ht}$ is sufficiently close to $T_\mathrm{st}$, the resulting locking into limit cycles is robust.

\begin{figure*}[t]
    \centering
    \includegraphics[width=1\textwidth]{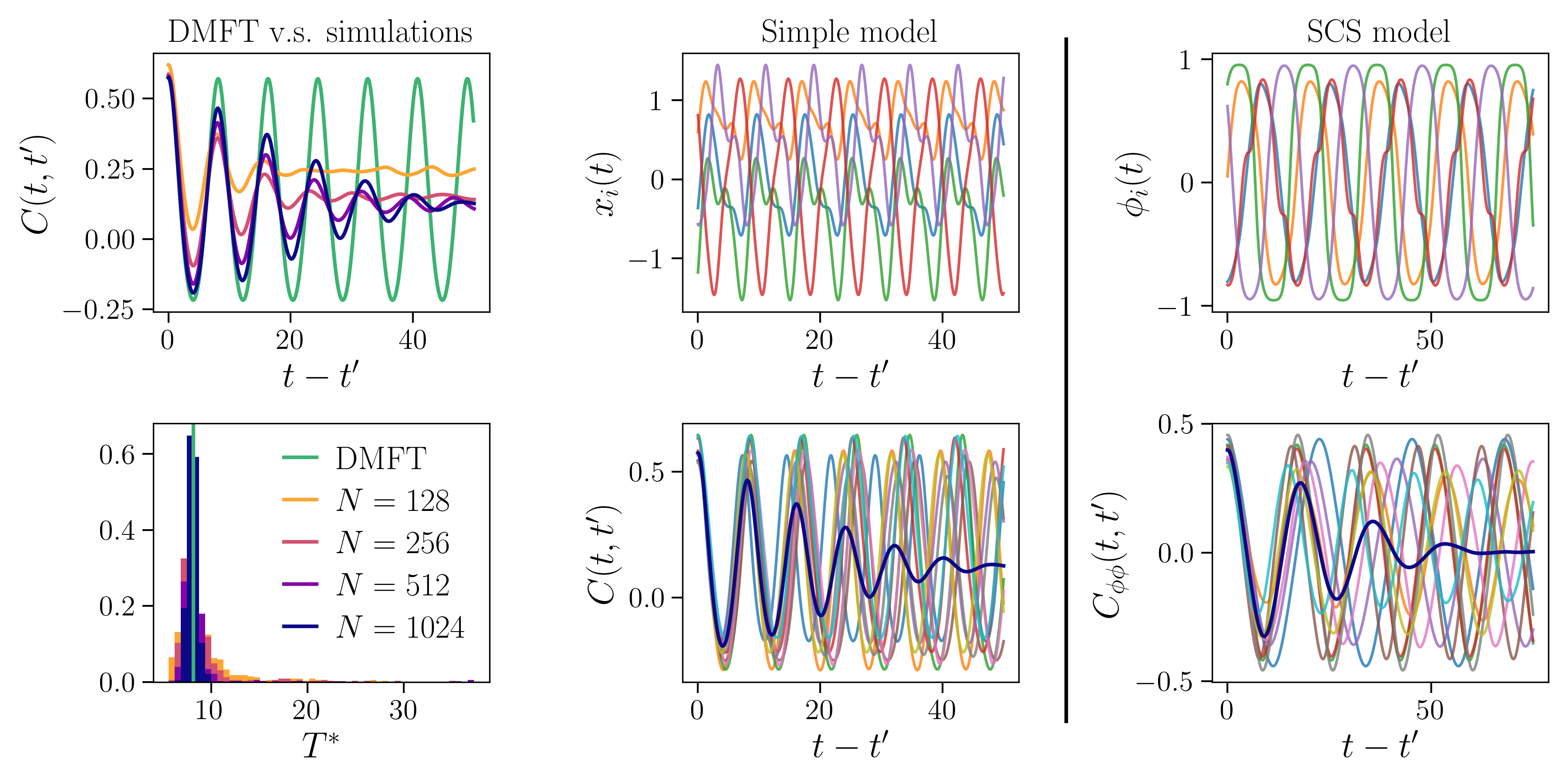}
    \caption{Comparison between DMFT and finite-size numerical simulations in the memorization phase. The left and middle panels show results for the dynamical system Eq.~\eqref{dyn_sys}, while the right panel is obtained with the SCS model. $C(t,t')$ as obtained from numerical simulations with $N$ neurons and averaged over $N_\mathrm{s}$ disorder realizations converges slowly as $N$ increases to its DMFT prediction (Top-left). This is due to sample-to-sample fluctuations in the period $T^*$ of the oscillations in $C(t,t')$, which result in destructive interference (Bottom-left). (Middle) Numerical simulations with $N=1024$ neurons in the memorization phase $t\geq t'\gg T_\mathrm{ht}$ with $I=2$, $k=2.5$. Traces of neuronal activities $x_i(t)$ (top) and sample-to-sample behavior of $C(t,t')$ (bottom) are shown. The dark curve is averaged over $N_\mathrm{s}$ realizations of disorder. (Right) Same as middle panels but for $\phi_i(t)$ and $C_{\phi\phi}(t,t') = \frac1N \sum_i \phi_i(t) \phi_i(t')$ in the SCS model. Parameters for the SCS model are: $g=1.5$, $p=15$, $\omega=0.6$, $I=2$, $k=3$.
    For both models, $N_s=1000$.}
    \label{fig:simulations}
\end{figure*}

\paragraph*{\bf Dynamical order parameters --}
The dynamical system in Eq.~\eqref{dyn_sys} can be analyzed in the $N\to \infty$ limit via dynamical mean field theory (DMFT) following the same approach developed in~\cite{fournier2023statistical,fournier2026high}, see the end matter (EM). DMFT allows to track the dynamics of a few coarse-grained observables. In particular, we can track the dynamics of the two-time correlation function
\begin{equation}
    C(t,t') = \lim_{N\to \infty}\frac{\<\bx(t),\bx(t')\>}{N}\:.
\end{equation}
This quantity allows also to obtain the Mean Square Displacement (MSD)
\begin{equation}
\begin{split}
    Y(t,t')&=\lim_{N\to \infty}\frac{1}{N}|\bx(t)-\bx(t')|^2\\
    &=C(t,t)+C(t',t') -2C(t,t')
\end{split}
\end{equation}
which denotes the square euclidean distance between the state of the system at two times $t$ and $t'$.

\paragraph*{\bf Phase diagram --}
The two time functions $C$ and $Y$ allow to compute the phase diagram. This is showed in Fig.~\ref{fig:phase_diagram}. We focus on a specific realization of the model for which $G(C) = g_1^2 C + 2g_2^2 C^2$, and $\hat \mu(z) = 1 + z + z^2$.
The coupling constants are set to $g = 0.8$, $g_1 = 2$, and $g_2 = 1$. This specific choice ensures that the underlying system (with neither drive nor plasticity) has a chaotic attractor. The external periodic drive acts at a constant angular frequency and we fix it to $\omega = 1.5$, while the timescale of the Hebbian synaptic plasticity is fixed to $p = 15$. Finally, we set $T_\mathrm{st}=T_\mathrm{ht}$ and we will later comment on what happens when a metabolic phase is present.

The phase diagram is plotted in Fig.~\ref{fig:phase_diagram} and it is explored by varying the driving amplitude $I$ and the plasticity coupling strength $k$. 
We extracted it by numerically integrating the DMFT equations presented in the EM.

The red dashed line defines the phase diagram of the model in the stationary phase for $t<T_\mathrm{st}$. Above that line, the stationary phase is a limit cycle, while below the line, the asymptotic attractor is instead a chaotic phase with some remnants of the periodic input visible in the power spectrum of $C(t,t')$ seen as a function of $t-t'$ for $T_\mathrm{st}> t\geq t'\gg 1$. This phenomenology extends the one of~\cite{rajan2010stimulus} to the case in which plasticity is added to the equations of motion of the dynamical system.

We now turn to the fate of the dynamical system in the memorization phase ($t\gg T_\mathrm{ht}$). This is described by the colored regions of the phase diagram  that correspond to different dynamical attractors.

In the gray area, the system lands on a chaotic attractor characterized by $\lim_{t\to \infty}C(t,t)=C_0>0$ and $C_{\infty}=\lim_{t'\to \infty}\lim_{t-t'\to \infty}C(t,t') = C_{\infty}=0$.
One can show via the techniques developed in~\cite{fournier2026high} that the corresponding maximal Lyapunov exponent is positive in the whole region. Given that the dynamics is chaotic, the shape of $Y(t,t')$ for $1\ll t,\, t'$ is featureless and in particular one has a well defined limit $Y_{\infty}=\lim_{t'\to \infty}\lim_{t-t'\to \infty}Y(t,t') = Y_{\infty}>0$ which coincides with $Y_{\infty}=2C_0$.

The green region instead correspond to a fixed point phase: the dynamics land on a fixed point as $C_0=C_\infty$ and $Y(t,t')=0$ for $t,\, t'\gg T_\mathrm{st}$.

Between the green and gray area, one can have the formation of a pink island which correspond to a semi-freezable chaotic phase~\cite{clark2024theory, fournier2023statistical}: the dynamics lands on a chaotic attractor which is correlated with the last configuration the system visited before halting plasticity and drive. It follows that $0<C_\infty<C_0$. The limiting value of the MSD is therefore given by $Y_{\infty}=2(C_0-C_\infty)$.
Overall, these three phases extend the ones found in~\cite{clark2024theory,fournier2023statistical} to the case where a periodic forcing is added prior to halting synaptic plasticity.

For large enough $I$ and $k$, one has two additional phases. Increasing $k$ at fixed $I$, one can first encounter a transition to a semi-periodic chaotic phase. In this case, the dynamics lands on a chaotic attractor characterized by a positive maximal Lyapunov exponent. The power spectrum of $C(t,t')$ as a function of $t-t'$ has a remnant of the periodic drive. 

Increasing further the plasticity strength, one lands on a limit cycle phase where, after halting the synaptic plasticity, the trajectory of the system is a limit cycle. This is reflected in $Y(t,t')$, which periodically vanishes (up to $10^{-3}$ precision and on the timescales we have access to) for sufficiently large $t'$. After a period $T^*$, the dynamics comes back to the configuration visited at $t'$. The period of the limit cycle is not exactly the same as the one of the periodic forcing during the training phase.

\paragraph*{\bf Numerical simulations --}
The results of the DMFT analysis can be compared with finite-size numerical simulations, as shown in Fig.~\ref{fig:simulations}.
We focus on the limit cycle phase, where a periodic attractor emerges during the memorization phase.

For \( t > T_\mathrm{ht} \), each simulated sample settles into a limit cycle. However, in any finite system, the periods of these limit cycles $T^*$ exhibit sample-to-sample fluctuations. While the average period matches that of the infinite-system attractor described by DMFT, there remains a finite dispersion for any finite \( N \). The difference in oscillation frequencies generally leads to destructive interference after a timescale inversely proportional to the frequency dispersion (as per the Heisenberg theorem). Consequently, detecting the limit cycle phase through averaged numerical simulations is challenging.

\paragraph*{\bf The role of the metabolic phase --}
When $ T_\mathrm{st} < T_\mathrm{ht} $, synaptic plasticity has a time window during which it operates without periodic input. We call this time window the metabolic time $\Delta_\mathrm{me}=T_\mathrm{ht}-T_\mathrm{st}$. 
Recently, in~\cite{wakhloo2025associative}, the case where $ T_\mathrm{ht} \to \infty $ was studied via numerical simulations in the SCS model. The conclusion is that, for sufficiently large $ I $ and $ k $, the system locks into a limit cycle for $ t < T_\mathrm{st} $. However, for $ t > T_\mathrm{st} $, the dynamics exhibit a transient phase where oscillations are present but eventually decay. The lifetime $ T_\mathrm{lt} $ of these oscillations depends on the specific region of the phase diagram. We confirm this behavior in the model we study. Furthermore, we observe that as long as $ T_\mathrm{lt} > {\Delta_\mathrm{me}} $, the dynamical attractor reached during the memorization phase is a limit cycle, see the EM.

\paragraph*{\bf The SCS model --}
We studied whether the same phenomenology found in the dynamical system we investigated can be found in the standard SCS model. In this case the DMFT analysis is almost prohibitive~\cite{wakhloo2025associative}, which is the reason why the model we focus on is useful. However, we performed numerical simulations and detected a region in the phase diagram of the model where the resulting picture is very similar to the data presented in Fig.~\ref{fig:simulations}. The conclusion is that limit cycles can be learned via synaptic plasticity if the dynamical system is driven sufficiently strongly and plasticity is sufficiently strong. When we consider the possibility of having $T_\mathrm{st}<T_\mathrm{ht}$, we find that as soon as $T_\mathrm{ht}<T_\mathrm{lt}$, the model locks in a limit cycle. All these numerical findings suggest that the phenomenology we described in the dynamical systems in Eq.~\eqref{dyn_sys} is robust.

\paragraph*{\bf Discussion --}
We showed that Hebbian plasticity can stabilize limit cycle attractors in high-dimensional non-linear dynamical systems.
The dependence of the limit cycle phase on the timescale and strength of synaptic plasticity, $p$ and $k$, on the properties of the periodic stimulus, $\omega$ and $I$, as well as on the metabolic time window $\Delta_\mathrm{me}$, remains to be fully understood. From the perspective of learning to autonomously generate periodic patterns, as in FORCE learning~\cite{sussillo2009generating}, the objective is instead to learn a precise limit cycle with a predefined frequency and amplitude. This raises the interesting possibility of implementing a biologically plausible learning strategy that leverages the memorization capacity of Hebbian synaptic plasticity characterized in this Letter.

While we mostly focused on a single frequency stimulus, it would be interesting to generalize the analysis to stimuli having more complex structure. Without Hebbian plasticity, this has been studied in~\cite{fournier2026high}. 
This is particularly important in a context in which the learning protocol we used is viewed from a stimulation-recall perspective. If plasticity is turned on in the stimulation phase, the memorization (inference) phase could be used to make predictions. This is very similar to test-time training in LLMs~\cite{sun2020test} and indeed requires faster synaptic updates that may be present in the Hippocampus.

\paragraph*{\bf Acknowledgements --} PU acknowledges funding by the French government under the France 2030 program (PhOM - Graduate School of Physics) with reference ANR-11-IDEX-0003.
This work was granted access to the CCRT High-Performance Computing (HPC) facility under the Grant CCRTYear-login awarded by the Fundamental Research Division (DRF) of CEA.

\bibliography{refs}

@article{Clark2025Multiregion,
  author  = {Clark, David G. and Beiran, Manuel},
  title   = {Structure of activity in multiregion recurrent neural networks},
  journal = {Proceedings of the National Academy of Sciences},
  year    = {2025},
  volume  = {122},
  number  = {10},
  pages   = {e2404039122},
  doi     = {10.1073/pnas.2404039122}
}

@article{Tsodyks2000Synchrony,
  author  = {Tsodyks, Misha and Uziel, Asher and Markram, Henry},
  title   = {Synchrony Generation in Recurrent Networks with Frequency-Dependent Synapses},
  journal = {The Journal of Neuroscience},
  year    = {2000},
  volume  = {20},
  number  = {1},
  pages   = {RC50},
  doi     = {10.1523/JNEUROSCI.20-01-j0003.2000},
  pmid    = {10627627}
}

@article{generativeRNN,
  title = {Generative modeling through internal high-dimensional chaotic activity},
  author = {Fournier, Samantha J. and Urbani, Pierfrancesco},
  journal = {Phys. Rev. E},
  volume = {111},
  issue = {4},
  pages = {045304},
  numpages = {7},
  year = {2025},
  month = {Apr},
  publisher = {American Physical Society},
  doi = {10.1103/PhysRevE.111.045304},
  url = {https://link.aps.org/doi/10.1103/PhysRevE.111.045304}
}

@article{laje2013robust,
  title={Robust timing and motor patterns by taming chaos in recurrent neural networks},
  author={Laje, Rodrigo and Buonomano, Dean V},
  journal={Nature Neuroscience},
  volume={16},
  number={7},
  pages={925--933},
  year={2013},
  publisher={Nature Publishing Group},
  doi={10.1038/nn.3405}
}

@article{rajan2016recurrent,
  title={Recurrent network models of sequence generation and memory},
  author={Rajan, Kanaka and Harvey, Christopher D and Tank, David W},
  journal={Neuron},
  volume={90},
  number={1},
  pages={128--142},
  year={2016},
  publisher={Elsevier},
  doi={10.1016/j.neuron.2016.02.009}
}

@article{Nicola2017FORCE,
  author  = {Nicola, Wilten and Clopath, Claudia},
  title   = {Supervised learning in spiking neural networks with {FORCE} training},
  journal = {Nature Communications},
  volume  = {8},
  number  = {1},
  pages   = {2208},
  year    = {2017},
  doi     = {10.1038/s41467-017-01827-3},
  url     = {https://doi.org/10.1038/s41467-017-01827-3}
}

@article{asabuki2025predictivealignment,
  title   = {Taming the chaos gently: a predictive alignment learning rule in recurrent neural networks},
  author  = {Asabuki, Toshitake and Clopath, Claudia},
  journal = {Nature Communications},
  volume  = {16},
  number  = {6784},
  year    = {2025},
  doi     = {10.1038/s41467-025-61309-9},
  publisher = {Springer Nature}
}

@article{depasquale2018full,
  title={full-FORCE: A target-based method for training recurrent networks},
  author={DePasquale, Brian and Cueva, Christopher J and Rajan, Kanaka and Escola, George S and Abbott, L F},
  journal={PLOS ONE},
  volume={13},
  number={2},
  pages={e0191527},
  year={2018},
  publisher={Public Library of Science},
  doi={10.1371/journal.pone.0191527}
}

@article{fournier2023statistical,
  title={Statistical physics of learning in high-dimensional chaotic systems},
  author={Fournier, Samantha J and Urbani, Pierfrancesco},
  journal={Journal of Statistical Mechanics: Theory and Experiment},
  volume={2023},
  number={11},
  pages={113301},
  year={2023},
  publisher={IOP Publishing}
}

@article{fournier2025non,
  title={Non-reciprocal interactions and high-dimensional chaos: comparing dynamics and statistics of equilibria in a solvable model},
  author={Fournier, Samantha J and Pacco, Alessandro and Ros, Valentina and Urbani, Pierfrancesco},
  journal={arXiv preprint arXiv:2503.20908},
  year={2025}
}

@article{rajan2010stimulus,
  title={Stimulus-dependent suppression of chaos in recurrent neural networks},
  author={Rajan, Kanaka and Abbott, LF and Sompolinsky, Haim},
  journal={Physical Review E—Statistical, Nonlinear, and Soft Matter Physics},
  volume={82},
  number={1},
  pages={011903},
  year={2010},
  publisher={APS}
}

@inproceedings{sun2020test,
  title={Test-time training with self-supervision for generalization under distribution shifts},
  author={Sun, Yu and Wang, Xiaolong and Liu, Zhuang and Miller, John and Efros, Alexei and Hardt, Moritz},
  booktitle={International conference on machine learning},
  pages={9229--9248},
  year={2020},
  organization={PMLR}
}

@article{wakhloo2025associative,
  title={Associative synaptic plasticity creates dynamic persistent activity},
  author={Wakhloo, Albert J and Clark, David G and Abbott, LF},
  journal={bioRxiv},
  year={2025}
}

@article{clark2024theory,
  title={Theory of coupled neuronal-synaptic dynamics},
  author={Clark, David G and Abbott, LF},
  journal={Physical Review X},
  volume={14},
  number={2},
  pages={021001},
  year={2024},
  publisher={APS}
}

@article{urbani2026theory,
  title={Theory of learning of high-dimensional controlled non-linear dynamical systems (I): models and methods},
  author={Urbani, Pierfrancesco},
  journal={arXiv preprint arXiv:2606.07247},
  year={2026}
}

@article{fournier2026chaos,
  title={Chaos in high-dimensional dynamical systems with tunable non-reciprocity},
  author={Fournier, Samantha and Urbani, Pierfrancesco},
  journal={arXiv preprint arXiv:2601.04702},
  year={2026}
}

@article{fournier2026high,
  title={High-dimensional dynamical systems: co-existence of attractors, phase transitions, maximal Lyapunov exponent and response to periodic drive},
  author={Fournier, Samantha J and Urbani, Pierfrancesco},
  journal={Journal of Statistical Mechanics: Theory and Experiment},
  volume={2026},
  number={4},
  pages={043302},
  year={2026},
  publisher={IOP Publishing}
}

@article{sussillo2009generating,
  title={Generating coherent patterns of activity from chaotic neural networks},
  author={Sussillo, David and Abbott, Larry F},
  journal={Neuron},
  volume={63},
  number={4},
  pages={544--557},
  year={2009},
  publisher={Elsevier}
}

@article{sompolinsky1988chaos,
  title={Chaos in random neural networks},
  author={Sompolinsky, Haim and Crisanti, Andrea and Sommers, Hans-Jurgen},
  journal={Physical review letters},
  volume={61},
  number={3},
  pages={259},
  year={1988},
  publisher={APS}
}

\widetext

\section*{End Matter}

\section{Dynamical mean field theory}
In this section, we derive the dynamical mean field theory (DMFT) equations that allow us to study the behavior of the dynamical system Eq.~\eqref{dyn_sys} in the thermodynamic limit. We will only focus on the main steps of the derivation given that the whole methodology is already described in the literature. In particular, for the class of models that we study, the interested reader can look at \cite{fournier2026high}. 

We start by solving the dynamics of the Hebbian plasticity term which can be formally written as
\begin{equation}
    A_{ij}(t) =\frac 1p \int_0^t 
    \de s\, \hat k(s) e^{-(t-s)/p}\frac 1N x_i(s)x_j(s)
\end{equation}
where 
\begin{equation}
    \hat k(t) = \begin{cases}
    k & t\leq T_\mathrm{ht}\\
    0 & t> T_\mathrm{ht}
    \end{cases}
    \:.
\end{equation}
Plugging this expression into the dynamical system we get
\begin{equation}
    \dot x_i(t) = -\mu(t)x_i(t) + g\, r_i(\bx(t)) + \frac 1p \int_0^t\de s\, \hat k(s)e^{-(t-s)/p}C(t,s)x_i(s) + h_i(t)
\end{equation}
where $C(t,s)=\bx(t)\cdot \bx(s)/N$.
We then decompose the degrees of freedom $\bx$ as
\begin{equation}
    \bx(t) =\bx^{(0)}(t)+\bx^{(1)}(t)
\end{equation}
where the two components obey the following equations
\begin{equation}
\begin{split}
    \dot x_i^{(0)}(t) &= -\mu(t)x_i^{(0)}(t) +  \frac 1p \int_0^{t}\de s\, \hat k(s)e^{-(t-s)/p}C(t,s)x_i^{(0)}(s) + h_i(t)\\
    \dot x_i^{(1)}(t) &= -\mu(t)x_i^{(1)}(t) + g\, r_i(\bx(t)) + \frac 1p \int_0^t\de s\, \hat k(s)e^{-(t-s)/p}C(t,s)x_i^{(1)}(s) \:.
\end{split}
\end{equation}
The initialization is such that $\bx^{(0)}(0)=0$ so that $\bx(0)=\bx^{(1)}(0)$.

Following the same steps as in~\cite{fournier2026high}, we can reduce the dynamics of this high-dimensional non-linear system to the one of two degrees of freedom following a self consistent stochastic process given by
\begin{equation}
    \begin{split}
        \dot x^{(0)}(t) &= -\mu(t)x^{(0)}(t) +  \frac 1p \int_0^t\de s\, \hat k(s)e^{-(t-s)/p}C(t,s)x^{(0)}(s) + \hat I\cos(\omega t+\theta)\theta_H(T_{tr}-t)\\
        \dot x^{(1)}(t) &= -\mu(t)x^{(1)}(t) + g\, \eta(t) + \frac 1p \int_0^t\de s\, \hat k(s)e^{-(t-s)/p}C(t,s)x^{(1)}(s)
    \end{split}
\end{equation}
with $\theta_H$ the Heaviside theta function and
\begin{equation}
    \begin{split}
        \theta&\sim \Unif(0,2\pi)\\
        \E[\eta]&=0\\
        \E[\eta(t)\eta(s)]&=G(C(t,s))\:.
    \end{split}
\end{equation}
The two stochastic processes for $x^{(0)}$ and $x^{(1)}$ are uncorrelated.
Therefore, we have that
\begin{equation}
    \begin{split}
        C(t,s)&=C_0(t,s)+C_1(t,s)\\
        C_0(t,s)&=\langle x^{(0)}(t)x^{(0)}(s)\rangle\\
        C_1(t,s)&=\langle x^{(1)}(t)x^{(1)}(s)\rangle
    \end{split}
\end{equation}
where the brackets indicate either the average over the random phase $\theta$ or the average over the effective noise $\eta$.

We now proceed to deriving the equations for $C_0$ and $C_1$. 
We have that
\begin{equation}
    \partial_t C_0(t,t') = -\mu(t)C_0(t,t') + \frac 1p \int_0^t\de s\, \hat k(s)e^{-(t-s)/p}C(t,s)C_0(s,t') + L(t,t')
\end{equation}
where $L(t,t')= I\langle\cos(\omega t+\theta)x^{(0)}(t') \rangle \theta_H(T_\mathrm{st}-t)$ obeys the following equation
\begin{equation}
    \begin{split}
        \partial_{t'}L(t,t') &= -\mu(t')L(t,t') + \frac 1p \theta_H(T_\mathrm{st}-t)\int_0^{t'}\de s\, \hat k(s)e^{-(t-s)/p}C(t',s)L(t,s) + Z(t,t')\\
        L(t,0)&=0\ \ \ \  \forall t\\
        Z(t,t') & = I^2\langle\cos(\omega t+\theta)\cos(\omega t'+\theta) \rangle\theta_H(T_\mathrm{st}-t)\theta_H(T_\mathrm{st}-t') \\
        &= \frac{\hat I^2}2\cos(\omega(t-t'))\theta_H(T_\mathrm{st}-t)\theta_H(T_\mathrm{st}-t')
    \end{split}
\end{equation}
Note that, as soon as either $t$ or $t'$ are larger than $T_{st}$, we get $L=0$.
Furthermore, we have
\begin{equation}
\begin{split}
    \partial_tC_1(t,t') &= -\mu(t)C_1(t,t') + \frac kp \int_0^t\de s\, e^{-(t-s)/p}C(t,s)C_1(s,t')  + \int_0^{t'}\de s\, G(C(t,s))R_1(t',s) \\
    \frac 12 \frac{\de C_1(t,t)}{\de t} &=  -\mu(t)C_1(t,t) + \frac kp \int_0^t\de s\, e^{-(t-s)/p}C(t,s)C_1(s,t)  + \int_0^{t'}\de s\, G(C(t,s))R_1(t,s)\\
    \partial_tR_1(t,t') &= -\mu(t)R_1(t,t') + \frac kp \int_{t'}^t\de s\, e^{-(t-s)/p}C(t,s)R_1(s,t') +\delta(t-t')\:.
\end{split}
\end{equation}
These dynamical equations can be integrated numerically, see~\cite{fournier2026high} for a  discussion of this point. Indeed, they have a causal structure that allows to track the dynamics easily once a careful discretization of the derivative terms and integrals is chosen. 

\begin{figure*}[t]
    \centering
    \includegraphics[width=1\textwidth]{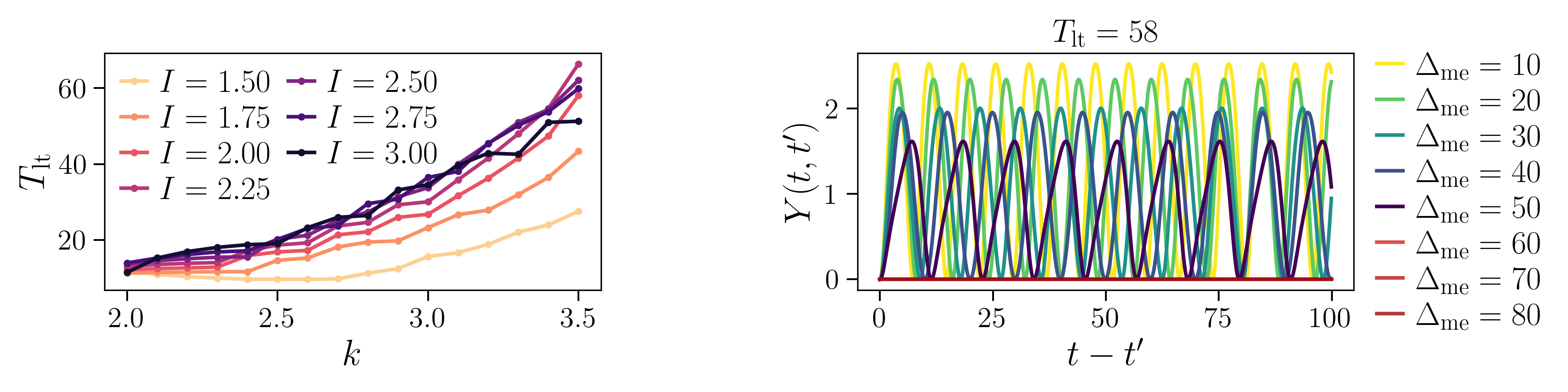}
    \caption{(Left) Lifetime of persistent oscillations for $t\geq T_\mathrm{st}$ with active plasticity ($T_\mathrm{ht}\to\infty$). The lifetime $T_\mathrm{lt}$ is defined as the time interval over which $C(t,T_\mathrm{st})$, plotted as a function of $t-T_\mathrm{st}$, displays sustained oscillations. Numerically, $T_\mathrm{lt}$ is obtained by locating the point after the final oscillation peak where $|\partial_t C(t,T_\mathrm{st}) |$ falls below a fixed threshold. (Right) Impact of a non-zero metabolic phase $\Delta_\mathrm{me}=T_\mathrm{ht}-T_\mathrm{st}$ on the limit cycle attractor of the memorization phase ($t\geq t' \gg T_\mathrm{ht}$). As long as $\Delta_\mathrm{me}$ is sufficiently small compared to $T_\mathrm{lt}$, the MSD vanishes periodically, indicating limit cycle activity (shades of viridis). Parameters are: $I=2$, $k=3.5$.}
    \label{fig:metabolic}
\end{figure*}

\section{Robustness of the limit cycle phase across a finite metabolic phase}
We explore how the scenario we have found for $T_\mathrm{st}=T_\mathrm{ht}$ is robust for $T_\mathrm{st}<T_\mathrm{ht}$.
We first consider the situation in which $T_\mathrm{ht}\to \infty$ and $T_\mathrm{st}$ is large but finite (so that during the training phase the dynamics leads to a stationary state).
It was shown in numerical simulations in~\cite{wakhloo2025associative} that, right after the periodic input is switched off, the dynamics of the system can have persistent oscillations. The period and lifetime of these oscillations is a function of the control parameter of the model. 
In the context of our model, we find the same phenomenology.
In Fig.~\ref{fig:metabolic}-left we plot the lifetime of the persistent oscillations for $t>T_\mathrm{st}$ as a function of the strength of the periodic input applied in the stimulation phase. This shows that there exist a region in the $I$-$k$ plane where the lifetime of the oscillations is long. 

We then pick a point in this phase diagram for which the lifetime $T_\mathrm{lt}$ of the oscillations is finite and consider the case in which $T_\mathrm{ht}$ is finite as well. The resulting metabolic window is $\Delta_\mathrm{me}=T_\mathrm{ht}-T_\mathrm{st}$. In Fig.~\ref{fig:metabolic}-right, we show the mean square displacement (MSD) of the dynamics $Y(t,t')$ for $t,\, t'\gg T_\mathrm{ht}$ large enough so that the system is in a stationary state. For $\Delta_\mathrm{me}<T_\mathrm{lt}$, the dynamics locks in a limit cycle, while for $\Delta_\mathrm{me}>T_{lt}$ the limit cycle is lost and for the precise point in the phase diagram considered in Fig.~\ref{fig:metabolic}, the dynamics lands on a fixed point.

\end{document}